\documentclass[10pt,conference]{IEEEtran}

\newcommand{\comment}[1]{}

\IEEEoverridecommandlockouts                  

\usepackage[T1]{fontenc}
\usepackage[utf8]{inputenc}
\usepackage{graphicx}
\usepackage{amsmath,amssymb}
\usepackage{booktabs}      
\usepackage{makecell}
\usepackage{multirow}
\usepackage{xcolor}
\usepackage{listings}      
\usepackage{url}
\usepackage[hidelinks]{hyperref}
\usepackage{cleveref}      
\usepackage{tikz}
\usetikzlibrary{arrows.meta, positioning, calc, shapes.geometric}

\begin{document}

\title{	Automated Testing and Repair for Verified Compilers Generated by a Coding Agent}


\author{
\IEEEauthorblockN{Martin Rinard \IEEEauthorrefmark{1}}
\IEEEauthorblockA{\IEEEauthorrefmark{1}National University of Singapore and Massachusetts Institute of Technology}
}

\maketitle

\begin{abstract}
We present an 
agent based automated testing and repair system for verified 
compilers that contain four kinds of code: verified code, checked code, unverified code, and specification. 
We present specialized defect detection techniques that exploit the structure present
in such compilers. 
For each surfaced defect the system invokes a coding agent to repair the defect and validate the repair. 
We evaluate the system on the Axon compiler, a compiler completely generated by a coding agent operating under developer supervision. The compiler was validated during development on a small benchmark set (the Livermore benchmarks), raising the possibility that its coding agent reward hacked the compiler. We also evaluate the possibility
that the repair system reward hacked the repairs and find no evidence of reward hacking
in either the Axon compiler or the repairs. 
We present results that characterize the testing and repair effectiveness and discuss repair characteristics.
\end{abstract}

\comment{
\begin{IEEEkeywords}
software engineering, keyword two, keyword three, keyword four
\end{IEEEkeywords}
}

\section{Introduction}
\label{sec:introduction}

Compilers play a central role in essentially all software development efforts. 
Because silent miscompiles break the language model that developers
use when reasoning about their software, compiler correctness is an important
property for effective software engineering. In response to this concern,
researchers have developed a range of testing techniques that specifically 
target compilers~\cite{chen2020survey,DBLP:journals/pacmpl/LivinskiiBR20,DBLP:conf/pldi/YangCER11,DBLP:conf/pldi/LeAS14,DBLP:conf/oopsla/LeSS15,DBLP:conf/oopsla/SunLS16,kang2018crellvm,DBLP:conf/pldi/LopesLHLR21}, with the surfaced defects then 
reported back  to the compiler developers to be repaired. An alternative approach
uses verified and/or checked 
compilation to ensure correct compilation~\cite{leroy2009backend,appel2014plcc,tan2019cakeml-jfp,rinard1999credible,rinard1999credible-cc,marinov2000,kang2018crellvm,zuck2002voc-conf,zuck2005validating,DBLP:conf/pldi/LopesLHLR21}.
All of these previous techniques work with compilers developed 
and maintained by human developers, with any defects manually repaired by those developers.

\noindent\textbf{ACDC.}
We present ACDC (Automated Certificate Detection and Correction), 
a new system and technique for automatically discovering and repairing defects 
in verified compiler systems that use credible/checked compilation. We apply
ACDC to Axon~\cite{Axon}, the first verified source to assembler compiler developed 
entirely by a coding agent operating under human supervision. Axon implements both verified
compiler techniques: 1) full verification of basic translation steps to ensure 
correct compilation for all programs and 2) checked credible compilation to ensure the compilation includes only
sound optimizations. With credible compilation optimizations are not verified ---
they instead generate a certificate that proves they transformed the program correctly. 
A verified certificate checker preserves correct compilation by rejecting incorrect
certificates. Axon also includes unverified text interfaces (parser and assembly language
printer) and an unverified assembly language operational semantics. 

ACDC exploits the structure of Axon to implement testing techniques that  
leverage Axon components, specifically certificates, the certificate checker, and 
formal operational semantics, that are not present in traditional unverified compilers. 
Working with these components, these testing techniques surface information 
that ACDC then presents to repair agents, which can 
use the information to precisely identify, 
localize, and repair defects or anomalies. 
For example:
\begin{itemize}
\item {\bf Certificates:} The Axon certificate checker checks 31 properties. If the certificate
fails to satisfy any of these properties, it rejects the certificate and produces a reject
record that identifies the optimization that produced the rejected certificate and the 
properties that the certificate failed to satisfy. When provided to the repair agent, 
this information enables the repair agent to identify the root cause for the rejection
and generate a repair that eliminates the rejection. Such repairs can target
the optimization, the optimization's certificate generator, or the certificate checker itself. 

If the repair updates the certificate checker, it must also update the certificate checker's
proof of correctness to reestablish verified end to end compiler correctness. 

\item {\bf Machine Model:} Axon works with a model of the underlying machine in the form
of a formal operational semanatics of the machine's assembly language instructions. 
Because the correctness proofs hold over this semantics, any deviation between the
semantics and the code as executed on the machine can cause a miscompilation. 
Executing an instruction sequence via the operational semantics and on the machine
makes it possible to identify the specific instruction within the sequence that
caused any discrepancy between the two, providing information that the repair agent can use
to identify a defective operational semantics rule and repair strategies for
that rule. 

The end to end proofs of compiler correctness are stated over this operational semantics.
Changes to the operational semantics therefore trigger proof updates to restore these
compiler correctness proofs. 

\end{itemize}

\noindent\textbf{Reward Hacking.}
All ACDC test scripts, harnesses, prompts, and repairs are written by a coding agent,  
not by human developers. Axon was also developed by a coding
agent, with a small set of benchmarks
(Livermore benchmarks~\cite{mcmahon1986livermore}) driving the development. ACDC's repairs
are driven by test cases that surface defects and certificate rejections. Coding
agents are known to reward hack generated code so that it does not generalize outside 
the test cases that drive its development and evaluation~\cite{DBLP:journals/corr/abs-2511-16858,DBLP:journals/corr/abs-2510-20270,DBLP:journals/corr/abs-2605-02964}. 
We use unseen randomized testing to evaluate whether Axon and ACDC reward hack
the code they generate:
\begin{itemize}
\item {\bf Axon:} Our randomized Axon testing indicates that the majority of 
randomly generated test inputs surface defects or certificate rejections. 
However, a small number of repairs (automatically
generated by the coding agent) eliminate all surfaced defects or rejections on newly
generated unseen random inputs. Our conclusion is that the Axon development left point
defects in the code base but there was no systematic reward hacking that impairs
the meaningful generalization of the system beyond the specific benchmark
set that drove the development. 

\item {\bf ACDC:} The fact that a modest number of repairs enable Axon to successfully
process subsequently generated random inputs indicates that the repairs themselves are
not systematically reward hacked. With several exceptions, the repairs update disjoint
pieces of code, consistent with each repair eliminating a point defect. 
We further examine the code for each repair and
conclude that the repairs are not reward hacked. 

\end{itemize}

\noindent\textbf{Research Questions.} We consider the following research questions:

\noindent{\bf RQ1:} What testing targets and techniques does the structure of the Axon verified compiler enable? We answer
this question in Section~\ref{sec:targets} by presenting these techniques. 

\noindent{\bf RQ2:} What information can these testing techniques provide to repair agents? We answer this
question in Section~\ref{sec:targets}, where we describe this information. 

\noindent{\bf RQ3:} Given this information, how effectively can the repair agents produce repairs? We
answer this question in Section~\ref{sec:results}, which presents experimental results that demonstrate
that the repair agents repaired every detected defect and
eliminated every detected certificate rejection.

\noindent{\bf RQ4:} How many resources did the tests and repairs consume and
which components, and how many files and lines of code, did the repairs 
modify? 
We answer this question in Section~\ref{sec:results}, which presents these numbers. 

\noindent{\bf RQ5:} Both Axon and ACDC were developed entirely by coding agents. Do the tests and
repairs provide any evidence that the code that the agents generated was reward hacked? 
We answer this question in Section~\ref{sec:results}, which analyzes the
testing data and generated repairs. 

\noindent\textbf{Contributions.}
This paper makes the following contributions:
\begin{itemize}
\item {\bf Testing Techniques:} It presents testing techniques that leverage the structure
and components of the Axon compiler system to precisely target different defect classes
present in specific unverified components (the text interfaces, machine model, and 
optimizations). These techniques combine randomized test input generation, instrumented 
executions, and automated agent code audits to surface defects or certificate rejections, 
then generate records that contain information designed to enable 
repair agents to identify, localize, and repair the defect or rejection. 

\item {\bf Automated Agent Repair:} It presents the first automated agent repair system
for compilers. This system provides the repair agent with the defect or rejection records
from the testing system and a prompt that instructs the agent how to localize and repair 
the defect or rejection. 

\item {\bf Proof Updates for Program Repairs:} It presents the first system that
updates proofs of program correctness in response to source code changes
(including changes in verified source
code). Specifically, operational semantics updates require updates to the 
end to end compiler correctness proofs; certificate checker updates require
updates to the certificate checker correctness proof. 

\item {\bf Experimental Results:} It presents experimental results
that characterize the effectiveness of the presented techniques. These
results show that the generated tests quickly saturate, that the
autonomous agent repairs all identified defects and rejections, and 
that the repairs show no evidence that either Axon or the repairs
were reward hacked by the agent. 

\end{itemize}

As the capabilities of coding agents continue to improve, software will 
increasingly be generated by coding agents, with verification and checked
computations playing an important role in ensuring the integrity of the
generated code. This transition will trigger exploration of new testing and repair techniques 
tailored for systems that contain verified, checked, and unverified components. 
We hope that the techniques and results presented in this paper can help
guide the evolution of the field as it navigates this 
transition. 


\begin{figure*}[t]
\centering
\resizebox{\textwidth}{!}{%
\begin{tikzpicture}[
  >=Stealth,
  node distance=0.55cm and 0.7cm,
  font=\small\sffamily,
  procunv/.style = {  
    draw=red!60!black, line width=0.6pt,
    fill=red!12,
    ellipse,
    minimum width=2cm, minimum height=0.9cm,
    align=center, inner sep=1pt
  },
  procver/.style = {  
    draw=green!45!black, line width=0.6pt,
    fill=green!12,
    ellipse,
    minimum width=2cm, minimum height=0.9cm,
    align=center, inner sep=1pt
  },
  procchk/.style = {  
    draw=yellow!50!black, line width=0.6pt,
    fill=yellow!30,
    ellipse,
    minimum width=2cm, minimum height=0.9cm,
    align=center, inner sep=1pt
  },
  datatext/.style = {  
    draw=black!45, line width=0.6pt,
    fill=white,
    rounded corners=3pt,
    minimum width=1.3cm, minimum height=0.8cm,
    align=center, inner sep=3pt
  },
  dataIR/.style = {  
    draw=black!55, line width=0.6pt,
    fill=black!10,
    rounded corners=3pt,
    minimum width=1.3cm, minimum height=0.8cm,
    align=center, inner sep=3pt
  },
  opsem/.style = {
    line width=0.6pt,
    rounded corners=3pt,
    minimum width=3.6cm, minimum height=1.0cm,
    align=center, inner sep=3pt,
    font=\footnotesize\sffamily
  },
  opsemblue/.style  = {opsem, draw=blue!55!black,  fill=blue!8},
  opsemgreen/.style = {opsem, draw=green!45!black, fill=green!10},
  opsempink/.style  = {opsem, draw=red!60!black,   fill=red!12},  
  swatch/.style = {
    draw=black!60, line width=0.4pt,
    minimum width=0.32cm, minimum height=0.32cm,
    inner sep=0pt
  },
  klabel/.style = {anchor=west, font=\small\sffamily, inner sep=1pt},
  arr/.style = {->, thick, draw=black!70, >=Stealth}
]
\node[datatext]                    (textprog) {text\\program};
\node[procunv,  right=of textprog] (parser)   {parser};
\node[dataIR,   right=of parser]   (ast)      {AST};
\node[procver,  right=of ast]      (flatten)  {flattening};
\node[dataIR,   right=of flatten]  (tac)      {TAC};
\node[procver,  right=of tac]      (codegen)  {codegen};
\node[dataIR,   right=of codegen]  (asm)      {ASM};
\node[procunv,  right=of asm]      (pp)       {printer};
\node[datatext, right=of pp]       (textasm)  {text\\assembly};

\node[dataIR,   above=1.1cm of tac] (taccert) {TAC +\\certificate};
\node[procchk,  left=of taccert]    (opt)     {optimization\\pass};
\node[procver,  right=of taccert]   (checker) {certificate\\checker};

\node[opsemblue,  below=0.7cm of ast] (astsem) {AST\\Operational Semantics};
\node[opsemgreen, below=0.7cm of tac] (tacsem) {TAC\\Operational Semantics};
\node[opsempink,  below=0.7cm of asm] (asmsem) {ASM\\Operational Semantics};

\node[swatch, fill=green!12,  draw=green!45!black,  anchor=west]
     at ([yshift=2.1cm]textprog.north west) (kL1) {};
\node[klabel, right=0.1cm of kL1] {Verified};
\node[swatch, fill=yellow!30, draw=yellow!50!black, anchor=west, below=0.10cm of kL1] (kL2) {};
\node[klabel, right=0.1cm of kL2] {Checked};
\node[swatch, fill=red!12,    draw=red!60!black,    anchor=west, below=0.10cm of kL2] (kL3) {};
\node[klabel, right=0.1cm of kL3] {Unverified};

\node[swatch, fill=black!10, draw=black!55, anchor=west]
     at ([yshift=2.1cm, xshift=-3.5cm]textasm.north east) (kR1) {};
\node[klabel, right=0.1cm of kR1] {Intermediate Representation};
\node[swatch, fill=blue!8,   draw=blue!55!black, anchor=west, below=0.10cm of kR1] (kR2) {};
\node[klabel, right=0.1cm of kR2] {Specification};

\draw[arr] (textprog) -- (parser);
\draw[arr] (parser)   -- (ast);
\draw[arr] (ast)      -- (flatten);
\draw[arr] (flatten)  -- (tac);
\draw[arr] (tac)      -- (codegen);
\draw[arr] (codegen)  -- (asm);
\draw[arr] (asm)      -- (pp);
\draw[arr] (pp)       -- (textasm);

\draw[arr] (tac)     -- (opt);
\draw[arr] (opt)     -- (taccert);
\draw[arr] (taccert) -- (checker);
\draw[arr] (checker) -- (tac);

\draw[arr] (astsem) -- (ast);
\draw[arr] (tacsem) -- (tac);
\draw[arr] (asmsem) -- (asm);

\end{tikzpicture}%
}
\caption{Axon compiler structure, annotated with operational semantics and trust-category coloring.}
\label{fig:structure}
\end{figure*}
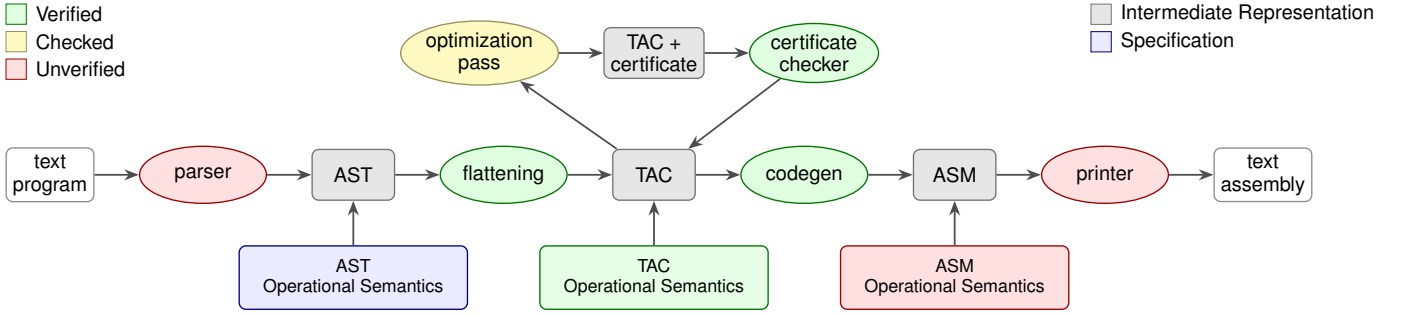

\section{Axon}

Figure~\ref{fig:structure} presents the structure of the Axon compiler. The verified 
core starts with a program represented as an abstract syntax tree (AST) implemented in 
Lean data structures. A flattening phase flattens nested expressions to 
produce a three address code (TAC) representation. A sequence of optimization
passes transforms the program. Each pass produces a transformed program plus 
a certificate proving that the transformed program {\em refines} the original program ---
i.e., that if the transformed program exhibits a given outcome (halts 
with given values in the observable variables, diverges, 
or divide by zero) then the original program can exhibit that same outcome. 
A certificate checker checks if the certificate establishes the refinement. 
If not, it rejects the transform and compilation continues with the next pass. A codegen
phase emits an assembly language (ASM) representation of the program implemented as
Lean data structures. The flattening phase, certificate checker, and codegen phase all
come with mechanically verified Lean refinement proofs --- 
the AST$\rightarrow$TAC, TAC$\rightarrow$ASM, 
and TAC$\rightarrow$TAC transitions accepted by the certificate checker are
all verified to satisfy refinement.  For clarity, the certificate checker comes with a proof
that if it does not reject the transformed program, then the transformed program 
refines the original program. 

To enable the verification, Axon includes an operational semantics for each program
representation (AST, TAC, and ASM) and uses these operational semantics to define
refinement. The AST operational semantics is the specification --- it defines
correct program behavior. The ASM operational semantics models the behavior of the
underlying machine that Axon compiles to. The TAC semantics bridges the AST and ASM
semantics and is therefore de facto verified by the end to end Axon correctness proofs. 

Axon also includes text interfaces --- a parser for a text representation of AST programs
and an ASM printer that generates text assembly files that can then be assembled,
linked, and executed. As is often the case for verified compilers~\cite{DBLP:conf/esop/MonniauxB22,DBLP:conf/tap/MonniauxGBL23}, the text interfaces 
(AST parser and ASM printer) are unverified. 

Note that (in the absence of a formal model of the underlying hardware) the ASM 
operational semantics is unverifiable --- the correctness
criteria would be that it accurately models the semantics of the actual hardware (for which 
a formal model is unavailable). In principle, the 
ASM operational semantics makes the combination
of the ASM operational semantics and the ASM printer testable --- generate
an ASM test program, compute its outcome via the operational semantics and via
generating and executing the text assembly file, and compare the outcomes. Any
difference surfaces a defect in either the operational semantics, the 
printer, or both. 

In practice, however, the Axon ASM operational semantics is structured as
a Lean proposition, which is not executable in Lean. To enable testing, we instructed
the coding agent to generate an executable version of the ASM operational semantics
and prove it equivalent to the nonexecutable version in the Axon compiler. The
experiments use this proved equivalent executable ASM operational semantics. 

\section{Certificates and the Certificate Checker}

Axon optimizations use credible compilation --- in addition to the transformed
program, they produce a certificate that proves the transformation preserved
the semantics.  Certificates and the verified certificate checker therefore
play a central role in the success of Axon. 

\subsection{Certificates} 

Optimizations prove they preserve semantics by generating a proof that the 
transformed program correctly simulates the original program. In addition
to a list of observable variables and a typing context that specifies a 
type (integer, float, or boolean) for each variable, the certificate
contains the following components:
\begin{itemize}
\item {\bf Single Program Invariants:} For each instruction in the original 
and transformed programs, invariants that hold at the instruction. Each invariant
equates a variable with an expression and states the invariant that, at that instruction, the
value of the variable equals the value of the expression. 

\item {\bf Relational Invariants:} For each instruction in the transformed program,
a corresponding instruction in the original program along with a list of pairs of 
equated expressions, one expression from the original program and the second
expression from the transformed program. The pair states the invariant that the
values of the two expressions at the two instructions (one from the 
transformed program and the corresponding instruction from the original program)
are equal, with each expression evaluated in the store of its corresponding program. 

\item {\bf Transitions:} At each instruction in the transformed program, control
may flow to one of several successor instructions. For each successor instruction, 
the certificate records a corresponding (potentially empty) {\em path} in the 
original program (the path is a 
sequence of instructions
in the original program). Each execution step (from one instruction to a successor
instruction) in the transformed program simulates executing the corresponding 
sequence of original program instructions. 

\end{itemize}

The transitions component, in effect, tells the
certificate checker how to structure the simulation proof that establishes that
the transformed program refines the original program. The single and relational
invariants establish the correspondence between variables and expressions in the
original and transformed programs, enabling the checker to match corresponding
conditional branch directions and observable variable values in the two programs. 

\subsection{Certificate Checker}

The certificate checker processes each certificate to check that it establishes the
required refinement relationship. It checks 31 conditions whose validity
together establishes the validity of the certificate. Relevant conditions 
triggered during testing runs include:
\begin{itemize}
\item {\bf all\_transitions:} False if one or more of the transitions specified in 
the certificate fails to check. The check requires 1) for each successor of
the instruction in the transformed program, the certificate specifies a corresponding
path in the original program, 2) the original program can walk the specified path, 
3) the effects agree --- a) all array writes match (array, index, value) and b) 
the relational invariant at the successor and last instruction in the path holds
after symbolically executing the instruction and path and substituting through 
the starting relational invariant. 

all\_transitions is by far the most common reason for certificate rejection --- 
any optimization pass that changes the control or data flow must generate
a certificate with nontrivial transitions around the optimization. 

\item {\bf invariants\_preserved:} False if one of the single program invariants
fails to verify. Checked at each instruction by symbolically executing the 
instruction against the instruction's invariant, simplifying, then checking 
if the resulting new invariant matches the invariant at the successor specified
in the certificate. 

\item {\bf div\_preservation:} False if the optimization may introduce a new 
divide by zero fault. Checked by requiring each divide in the transformed program
to map (via the transitions component) to a corresponding divide in the original
program and either 1) the two divide instructions have the same variable 
as the divisor and the variable has the same value in both programs or 
2) the divisor in the transformed program is a nonzero literal and the original 
program invariant proves that the divisor in the original program is nonzero. 

\item {\bf bounds\_preservation:} False if the optimization may introduce a new 
out of bounds array access fault. Checked by requiring each array access in the
transformed program to map (via the transitions component) to a corresponding 
array access such that 1) the two accesses access the same array with the same
access (load or store), 2) the array has the same size in the original and transformed
program, and 3) the array indices evaluate to the same value in both programs. 

\item {\bf bool\_vars\_covered:} Processes the branch condition for every conditional
branch instruction in the transformed program to check if every variable in the 
branch condition is mapped via the relational invariant to some expression in the
original program. The certificate checker proof of correctness requires that the 
checked conditions ensure corresponding flow of control at corresponding branches --- i.e., that
the value of the branch condition is the same in the transformed and original programs.
This condition ensures that the variable mappings exist to perform the reasoning required
to check corresponding flow of control. Note that this reasoning is performed once in the
certificate checker correctness proof, with bool\_vars\_covered discharging
a hypothesis in this proof. 

\item {\bf orig\_mismatch:} False if the original program in the certificate does not
match the original program passed to the optimization. Strictly speaking not 
a certificate validity issue but checked by the driver that invokes the sequence of
optimizations. 

\end{itemize}

\section{Testing and Repair Targets and Techniques (RQ1,RQ2)}
\label{sec:targets}

Axon (and other verified compilers) comes with components not available in 
standard unverified compilers --- these components include
operational semantics, certificates, and certificate checkers. These
components enable targeted testing and repair techniques not available to standard
unverified compilers. 

\subsection{ASM Operational Semantics and Printer}


The ASM operational semantics 
1) identifies all of the instructions that the compiler can ever emit and 2) 
provides an operational semantics for these machine instructions. 
The combination of the ASM operational semantics and printer 
therefore comprises a focused testing and repair target that 
is unavailable to standard compilers, which simply emit machine
code with no explicit semantics. 
The Axon correctness proofs are 
proved against the ASM operational semantics --- if the operational semantics
does not accurately reflect the semantics of the target machine, the compiled
program can generate the wrong outcome when run on the target machine.\footnote{
Although the headline correctness proofs guarantee that the ASM program execution 
is correct if executed by the executable ASM operational semantics.}

\noindent{\bf Testing:} 
The test script generates random ASM instruction sequences between 6 and
32 instructions in length. It generates ASM instruction sequences in batches, with each 
batch given a new random seed (the repair agent uses these seeds to reproduce the
instruction sequences during repair validation). The script uses 
the executable operational semantics to simulate the 
instruction sequence, invokes the printer to generate a text assembly file,
then assembles, links, and executes the file. Both executions start from
a common identical machine state. Constants such as integer immediates and 
floating point values are chosen randomly, either the random number
itself (probablity 1/3) or from a curated pool of values (extremal numbers, 
powers of two $\pm 1$, etc., probability 2/3). There are four potential outcomes:
\begin{itemize}
\item {\bf Match:} Both executions complete normally and the final machine
states match. No defect. 
\item{\bf Different Outcome:} Both executions complete normally, final machine
states differ. Defect. 
\item {\bf Assemble Fail:} The generated assembly file does not assemble. Defect.
\item {\bf Panic:} The executable operational semantics panics because of a
panic in either Lean or the Lean C runtime. Defect. 
\end{itemize}
For each surfaced defect, the script writes a defect record 
that identifies the defect class, the assembly language sequence that 
exposed the defect, and the random seed for the instruction sequence's batch.  
For Different Outcome defects the record also contains 
the registers whose values differ. For Assemble Fail defects the record also 
contains the error message from the assembler. 

\noindent{\bf Repair:} 
The repair agent is given a file of defect records and a prompt instructing
it to repair the defects that the file identifies. The prompt tells the 
agent how to reproduce the defect and that root cause defects can occur either in the 
ASM operational semantics or the ASM printer. The prompt instructs the
agent to proceed as follows: 1) identify a defect to fix, 2) audit the 
ASM operational semantics, every proof site that consumes the operational
semantics, and the agent's own understanding of each instruction
in the defect record, 3) generate a repair, 4) use the random seed in the 
defect record to reconstruct the batch that generated the instruction sequence,
and 5) validate the repair against the reconstructed batch and a new batch of
instruction sequences with a new seed. The agent loops until it finds a 
repair that validates. Code audits and validation steps can surface
new defects, with the new defects staged for later repair. 

\subsection{Optimizations, Certificates, and Certificate Checker}

Because of credible compilation, no Axon optimization can cause an incorrect
compilation --- if the optimization is incorrect, 
the checker will reject
the certificate and the optimization will be discarded. Indeed, an advocated
advantage of credible compilation is the ability to safely incorporate 
incorrect but still useful optimizers into the compiler~\cite{rinard1999credible}. 

\noindent{\bf Testing:} 
The testing goal is to identify certificate rejections, which
can happen for several reasons:
\begin{itemize}
\item {\bf Unsound Transform:} The tranformation itself is unsound --- there is 
no valid certificate for the transform.
\item {\bf Incorrect Certificate:} The transform is sound but the transform's
certificate generator produced an incorrect certificate for the transform. 
\item {\bf Tight Certificate Checker:} The transform and certificate are sound, but the
certificate checker rejects the certificate. Note that this situation may not require
a repair --- the range of valid certificates that the checker should accept is an 
engineering decision gated on factors that may include, for example, how often correct
certificates are rejected and the engineering effort required to generalize the checker
to accept the certificate. 
\end{itemize}
The test script repeatedly generates programs, then executes the standard Axon
optimization sequence on the program, with the certificate checker generating a 
rejection record for every certificate it rejects. This rejection record includes 
the name of the optimization pass whose certificate was rejected, the reason
for the certificate rejection (as above), and a program that triggers the 
certificate rejection. 

The test script uses three random test-generation variants, run in sequence so that each variant
tests the compiler as repaired by the previous variant. The first (T3a) uses a
Csmith~\cite{DBLP:conf/pldi/YangCER11} style type directed random program generator. It effectively
has a separate grammar for expressions of each type and randomly generates a sequence of statements
with randomly generated expressions. The second (T3b) applies Skeletal Program Enumeration
(SPE)~\cite{DBLP:conf/pldi/ZhangSS17} style techniques to seed programs generated by the T3a Csmith
style generator. SPE preserves the program structure but changes the dataflow by randomly scrambling
which variables occur at all program points of the same type; the implementation generates six SPE
variants per Csmith seed program. The third (T3c) applies Equivalence Modulo Inputs (EMI) style
techniques to seed programs generated by the T3a Csmith style generator. There are three variants
inspired by previous work: Orion~\cite{DBLP:conf/pldi/LeAS14} (delete dead code),
Athena~\cite{DBLP:conf/oopsla/LeSS15} (insert and delete code in dead regions), and
Hermes~\cite{DBLP:conf/oopsla/SunLS16} (insert code in live regions with guards that ensure the code
does not execute). All techniques preserve program semantics.

\noindent{\bf Repair:} 
The repair agent is given a file of rejection records and a prompt instructing
it to eliminate the rejections. The prompt informs the agent that it can modify
the transform, the certificate generator, the certificate checker, or any combination of the three. The prompt tells the 
agent to 1) reproduce the defect using the trigger program in the rejection record,
2) determine if the rejection was caused by an incorrect transform
or the certificate checker rejecting a correct certificate, 3) audit the code
for the optimization that caused the rejection and the code for the rejection 
reason (in the certificate generator) as well as every proof and code site that consumes
them, 4) localize the root cause, 5) make the minimal targeted repair that makes a
valid certificate accept, and 6) validate the repair on the trigger program and several
other generated programs. The prompt emphasizes that the agent should make sure that the
repair preserves the headline correctness theorems. In practice different repairs
target the transform, the certificate generator, and loosen the certificate
checker to accept more certificates (including repairing the certificate correctness
proofs required to preserve the headline correctness theorems). 

\subsection{Parser and AST Printer}

Axon comes with a parser and an AST printer. 
The goal is to find and repair parse
failures and mismatches between the two so that printing and parsing an AST
program produces the original program. Note that repairs can legitimately target
the parser, the printer, or both. 

\noindent{\bf Testing:} 
The test script randomly generates AST programs, 
prints them to text, parses them back to AST, and checks if the 
printed AST equals the generated AST. While the technique detects (and
repairs) parse failures, it is not designed to surface parser defects that 
occur only on malformed inputs --- 
it is instead geared toward ensuring the parser and printer are consistent. Like 
the ASM generator, it randomly generates AST programs in batches, with one
random seed per batch. Type directed recursive expression generators produce
random expressions, with random constants (probability 1/3) and constants randomly
selected from a pool of extremal constant values (probability 2/3). 
If the parse fails the script generates a defect record that records
the random seed, the error message, and the program that failed the parse. 
The script discards duplicate records with the same random seed and error message,
with a cap of 20 records with the same error message across all batches. If 
there is a mismatch between the generated and parsed AST programs, the 
script generates a defect record with the seed and the program that failed
the parse, again imposing a cap of 20 records across all batches. 

\begin{table*}[t]
  \centering
  \caption{Operational Semantics and ASM Printer Repairs. 
Detect Phase: D = found in detect phase; U\,(\textit{x}) = unmasked during verification of fix~\textit{x}. 
Time/tokens/cost/tool-calls are for the fix session that landed the repair; any earlier failed/escalated attempts are excluded.}
  \label{tab:t1-repairs}
  \setlength{\tabcolsep}{4pt}
  \footnotesize
  \begin{tabular}{@{}lll cc r rrrr r r@{}}
    \toprule
    Defect & \makecell{Detect\\Phase} & \makecell{Updated\\Component} & \makecell{Def LOC\\($+$/$-$)} & \makecell{Proof LOC\\($+$/$-$)} & \makecell{Time\\(min)} & \makecell{Input\\Tokens} & \makecell{Cache Read\\Tokens} & \makecell{Cache Write\\Tokens} & \makecell{Output\\Tokens} & Cost & \makecell{Tool\\Calls} \\
    \midrule
    Cmpimm & D & Printer & $+13/{-}3$ & $0/0$ & 9.8 & 4{,}032 & 2{,}425{,}712 & 58{,}427 & 28{,}688 & \$2.54 & 48 \\
    AndEor & D & Printer & $+25/{-}4$ & $0/0$ & 13.3 & 3{,}890 & 2{,}276{,}920 & 77{,}246 & 43{,}209 & \$3.01 & 38 \\
    Shift & D & Semantics & $+6/{-}6$ & $+2/{-}2$ & 14.2 & 4{,}725 & 2{,}844{,}945 & 69{,}049 & 32{,}153 & \$2.94 & 51 \\
    CsetCbnz & D & Printer & $+4/{-}4$ & $0/0$ & 12.2 & 8{,}269 & 2{,}381{,}734 & 67{,}901 & 27{,}535 & \$2.60 & 41 \\
    Fcmp & U\,(CsetCbnz) & Sem+Printer & $+21/{-}5$ & $+24/{-}24$ & 35.8 & 4{,}559 & 5{,}949{,}088 & 142{,}718 & 91{,}352 & \$6.71 & 67 \\
    \bottomrule
  \end{tabular}
\end{table*}

\noindent{\bf Repair:} The repair agent is given a file of defect 
records and a prompt instructing it to eliminate the defects. The prompt
instructs the agent to prioritize repairing the parser over the printer. 
It then instructs the agent to 1) group defect records by root cause, 2) choose 
a defect, 3) audit the relevant parser and printer code, 4) decide which is
wrong, 5) generate the repair, and 6) validate the repair 
against the batch that generated the detect record and a new batch of
programs generated with a new seed. If the validation surfaces new
defects it stages the defects for subsequent repair. 

\section{Experimental Results (RQ3,RQ4,RQ5)}
\label{sec:results}

We report experimental results applying our testing and repair techniques to the
Axon compiler. 

\subsection{Implementation and Methodoloy} 

We supervised the coding agent to implement the testing and repair techniques --- the
entire ACDC implementation, including all test scripts and all prompts, 
was automatically generated
by the coding agent. Each testing campaign
starts with a 30 minute (wall clock) detect phase that collects defect/rejection records
from runs on random inputs. In a following repair phase repair agents
then attempt to repair any surfaced defects or certificate rejections. 
The campaign then continues with another 30 minute detect phase that works
with the repaired system followed by another repair phase, iterating until the campaign converges. The operational semantics and parser campaigns converge when a single 30 minute detect phase surfaces no defects; the certificate checker campaigns converge when eight consecutive 30 minute detect phases (four hours) surface no certificate rejections. 
In all of our experiments the repair agents successfully repaired all of the surfaced
defects or rejections, including defects unmasked by repairs and surfaced during
repair validation. 
The operational semantics and parser campaigns generate tests sequentially; the certificate
checker campaigns run six parallel test generation processes in parallel. All run the 
repair agents sequentially, with each repair agent working on the updated code produced
by the previous repair agent. The operational semantics and parser campaigns run 
isolated in parallel starting from the baseline Axon compiler, with their repairs
consolidated as the starting point for the certificate checker campaign. 
All experiments were performed on a MacBook M4 Pro, 12 cores, 24 GB RAM, 
macOS 15.1 Darwin 24.1.0, ARM64 ISA with Claude Code Opus 4.8, VS Code
1.126.0, and Lean 4.28.0. Axon was developed on a MacBook M4 Pro 
with VS Code, Lean 4, and Claude Code Opus 4.6/4.7~\cite{Axon}.

\subsection{Operational Semantics and ASM Printer}

The initial 30 minute detect phase processes 8,200 assembly
programs, 82\% (6,760) of which trigger a defect.
Table~\ref{tab:t1-repairs} presents numbers from the ensuing repair phase.\footnote{
All numbers in these tables and presented in this section were computed by scripts generated by the
coding agent. These scripts process data collected by scripts (again generated by 
the coding agent) that execute during the detect and repair phases. 
Dollar costs are computed by applying published Anthropic Opus 4.8 API token rates
and do not reflect actual costs incurred by, for example, flat rate subscribers such as Claude Max subscribers.
Def LOC and Proof LOC in all repair tables count code lines only (Lean comments and blank lines excluded);
Proof LOC covers the propositional checker specification and the correctness proofs, Def LOC everything else.
}
The phase repaired five defects, four of which were surfaced during the detect phase (D) and
one of which was unmasked during defect repair validation (U). Three repairs targeted the Printer,
one targeted the Semantics, and one targeted both. The first four repairs were relatively small
and did not involve proof updates. The remaining statistics characterize
the resources required to perform the repairs, the first four of which landed quickly with small resource
consumption. The final repair (Fcmp) was the largest and the only one that required proof updates,
changing the semantics and printer ($+21/{-}5$ lines) together with the correctness proofs
($+24/{-}24$ lines). After the five repairs landed, the
repair agent was invoked a sixth time but surfaced no further defect (52 minutes, \$1.39), for a total
repair cost of \$19.19.
When the repair phase finished, the second detect phase processed 4,600 programs with no defects surfaced.

\subsubsection{Cmpimm}
The Cmpimm defect occurred when the printer processed register to immediate comparisons --- 
with the defect, large negative immediates
overflow the immediate field and the generated text fails to assemble. The repair updates
the Printer to generate code that correctly materializes large
immediates (negative or positive) into a scratch register and compares the registers.

\subsubsection{AndEor}
The AndEor defect occurred when the printer processed the bitwise \verb+andImm+/\verb+eorImm+
instructions --- it ignored the destination and source registers and emitted a hardcoded
32 bit \verb+and w0, w0, #imm+ (respectively \verb+eor w0, w0, #imm+), whereas the 64 bit
operational semantics writes the specified destination register reading the specified source
register (\verb+rd := rn & imm+, respectively \verb+rn ^ imm+). Because the emitted code writes
\verb+w0+ instead of the destination, any such instruction whose destination register is not the
scratch \verb+x0+ leaves that register unchanged on the machine while the model updates it, so the
model and machine end with different register values. The
repair updates the Printer to emit the actual registers, encoding the immediate as an AArch64 bitmask immediate when possible and otherwise materializing it into a scratch register.

\subsubsection{Shift}
The Shift defect occurs when a shift instruction attempts to shift by an amount larger
than 63. The executable operational semantics either panics Lean (very large immediates)
or produces a different value than the code executed on the machine (large immediates).
The repair changes the operational semantics to mod all shift amounts by 64.
No proof changes --- the proofs all go through verbatim with this change.

\subsubsection{CsetCbnz}
The CsetCbnz defect occurred when the Printer generated code for the conditional set (\verb+cset+) and
compare and branch nonzero (\verb+cbnz+) instructions. The printer hardcoded the scratch register
\verb+w0+, ignoring the specified operand register. For \verb+cset+ the result was written into
\verb+w0+ instead of the destination register; for \verb+cbnz+ the branch tested \verb+w0+
instead of the operand. As a result the printed code and the model can take different control flow paths --- for \verb+cbnz+, \verb+w0+ is always zero in the harness, so the printed code never branches while the model does --- and they end with different register values. The repair changed the printer to emit the
specified registers for both instructions.

\subsubsection{Fcmp}
The Fcmp defect occurred when the operational semantics stored the raw double bit patterns of a
float compare into the integer compare flags and then applied a signed-integer comparison. This is
wrong for IEEE-754 floating point numbers --- signed-integer order agrees with float order only for
positive values, so the comparison reverses the order of negative floating point numbers. It
also mishandles NaN (unordered) and $\pm 0$. The defect surfaced as a difference between the
operational semantics and the machine. The repair introduced an opaque \verb+fcmpFlags+ function
that decodes the IEEE four-way compare result and produces a canonical flags value
whose condition code evaluation produces the correct outcome for every condition and every value
class (less-than, greater-than, equal, and unordered/NaN). The integer flags structure and its
condition holds function are left unchanged, so the integer compare path and its proofs are
untouched. The repair also corrected the printer's float condition mapping to emit
unordered inclusive signed condition mnemonics, and restated the trusted float compare axiom over
\verb+fcmpFlags+. The proof updates ($+24/{-}24$ lines) reclose the semantics mirrors and the axiom
against the new flags. No headline theorem statement changed and no new axiom was introduced.

In general, the code surrounding these defects and repairs is well structured
and the repairs appear to generalize correctly across programs written in the language. The
exception is the proof code, which is extremely detailed but mechanically verified. 
We observed no evidence of reward hacking in either the Axon code or the repairs. 

\subsection{Transforms, Certificate Generators, and Certificate Checker}

The T3a (Csmith) campaign ran for 16 detect rounds. The first detect round produced certificate
rejections for 21 distinct combinations of optimization pass and failed checker property. The ensuing
repair phase produced eight repairs, seven for causes detected in that round (D1) and one (Unreachable)
unmasked during the round-one repair phase (U1). The second detect round produced three new
combinations, repaired by two fixes (D2). The third and fourth rounds each surfaced one new
combination (D3, D4), each repaired. Rounds five through seven produced no new rejections; the eighth
round surfaced one final combination (DAEstore, D8), repaired in the ensuing phase. The following eight
rounds (nine through sixteen) produced no new rejections, satisfying the convergence criterion of
eight consecutive clean 30-minute detect rounds. The T3a campaign tested 4,800 programs, surfaced 27
distinct rejection combinations, and produced 13 repairs. Table~\ref{tab:t3a-repairs} presents 
additional statistics for the T3a campaign certificate rejections and repairs. Six repairs targeted an optimization pass (its transform or certificate generator) and seven targeted the certificate checker; every checker repair also updated the checker's correctness proof.

\begin{table*}[t]
  \centering
  \caption{Certificate Checker Repairs (T3a, Csmith generator).
Detect Phase: D$n$ = detected in detect round~$n$; U$n$\,(\textit{x}) = unmasked in round~$n$ while fixing~\textit{x}. Component: optimization \emph{Pass} vs.\ certificate \emph{Checker}.
Time/tokens/cost/tool-calls are for the fix session that landed the repair.}
  \label{tab:t3a-repairs}
  \setlength{\tabcolsep}{4pt}
  \footnotesize
  \begin{tabular}{@{}lll cc r rrrr r r@{}}
    \toprule
    Defect & \makecell{Detect\\Phase} & \makecell{Updated\\Component} & \makecell{Def LOC\\($+$/$-$)} & \makecell{Proof LOC\\($+$/$-$)} & \makecell{Time\\(min)} & \makecell{Input\\Tokens} & \makecell{Cache Read\\Tokens} & \makecell{Cache Write\\Tokens} & \makecell{Output\\Tokens} & Cost & \makecell{Tool\\Calls} \\
    \midrule
    Compaction & D1 & Pass & $+1/{-}4$ & $0/0$ & 20.7 & 4{,}419 & 3{,}071{,}153 & 89{,}672 & 37{,}630 & \$3.40 & 51 \\
    Oscillation & D1 & Checker & $+8/{-}6$ & $+4/{-}1$ & 34.6 & 4{,}957 & 7{,}801{,}979 & 148{,}943 & 75{,}640 & \$7.31 & 81 \\
    DeadEdge & D1 & Checker & $+29/{-}11$ & $+133/{-}53$ & 69.7 & 6{,}541 & 22{,}095{,}930 & 258{,}716 & 159{,}000 & \$17.65 & 132 \\
    DivDividend & D1 & Checker & $+7/{-}2$ & $+29/{-}13$ & 42.0 & 9{,}618 & 11{,}415{,}398 & 172{,}589 & 98{,}804 & \$9.95 & 114 \\
    FMAtarget & D1 & Pass & $+2/{-}4$ & $0/0$ & 22.1 & 4{,}099 & 2{,}809{,}886 & 75{,}083 & 34{,}051 & \$3.03 & 47 \\
    Peephole & D1 & Checker & $+17/{-}16$ & $+143/{-}99$ & 52.6 & 5{,}284 & 17{,}595{,}367 & 270{,}514 & 176{,}299 & \$15.94 & 113 \\
    LICMpreheader & D1 & Pass & $+36/{-}1$ & $0/0$ & 43.3 & 8{,}479 & 9{,}588{,}469 & 188{,}708 & 122{,}996 & \$9.80 & 81 \\
    Unreachable & U1\,(DivDividend) & Checker & $+31/{-}0$ & $+196/{-}63$ & 79.2 & 15{,}337 & 65{,}958{,}942 & 418{,}547 & 251{,}984 & \$43.54 & 233 \\
    RegAlloc & D2 & Pass & $+10/{-}2$ & $0/0$ & 34.7 & 6{,}401 & 5{,}186{,}046 & 110{,}446 & 51{,}649 & \$5.02 & 60 \\
    CmpFold & D2 & Checker & $+4/{-}3$ & $+3/{-}1$ & 46.1 & 8{,}994 & 11{,}163{,}352 & 145{,}430 & 65{,}839 & \$8.73 & 107 \\
    BoolWrap & D3 & Checker & $+15/{-}11$ & $+15/{-}5$ & 41.8 & 6{,}246 & 8{,}790{,}333 & 131{,}258 & 44{,}159 & \$6.85 & 89 \\
    FMAtemp & D4 & Pass & $+3/{-}1$ & $0/0$ & 19.3 & 11{,}796 & 4{,}570{,}922 & 102{,}536 & 32{,}485 & \$4.18 & 54 \\
    DAEstore & D8 & Pass & $+1/{-}12$ & $0/0$ & 37.9 & 4{,}629 & 6{,}167{,}881 & 118{,}841 & 60{,}379 & \$5.81 & 66 \\
    \bottomrule
  \end{tabular}
\end{table*}

The T3b (SPE) campaign then ran against the compiler as repaired by T3a. Over nine detect rounds
(5,398 programs) it surfaced a single new certificate rejection, repaired by one checker fix, 
then produced eight consecutive clean rounds. Table~\ref{tab:t3b-repairs} presents statistics 
for the T3b campaign. The T3c (EMI) campaign, run against the compiler as repaired by T3a and T3b, 
surfaced no new rejections over eight detect rounds (3,758 programs).

\subsubsection{Compaction}
The ConstProp pass ran a compaction pass on both the original and the transformed program. This
phase removes unreachable code. The compiler driver rejects the certificate because the original
program in the certificate does not match the program given to the pass.  The repair drops the 
compaction step for both programs. Only the transform changes --- the certificate generator
and checker remain unmodified. We note that the downstream dead code elimination pass removes
unreachable code so that unreachable code removal is redundant in the constant propagation pass. 

\subsubsection{Oscillation}
The certificate checker normalizes a floating-point addition by moving a multiply operand to the
right. When both operands are multiplies, the rule merely swapped them, so the
checker's iterated simplifier oscillates and the two sides of an invariant check end up in opposite
operand orders. The result was a failing syntactic equality on a valid certificate. The repair makes the
normalization idempotent by leaving the operands unchanged when both are multiplies. 
The repair updated both the certificate checker and its correctness proof. It
left the transform and certificate generator unmodified.

\subsubsection{DeadEdge}
The certificate checker required a valid original program path for every syntactic successor of a
transformed conditional branch, including a branch edge that the invariant proves is never taken. 
The repair adds a dead branch test. It evaluates the branch condition under the original
invariant and skips the program path check for a provably dead successor. 
The repair updated both the certificate checker and its correctness proof. It
left the transform and certificate generator unmodified.

\subsubsection{DivDividend}
The divide preservation check in the certificate checker required the dividend of a divide to map to an
original program variable. A divide by zero fault, however, depends only on the divisor. When loop invariant
code motion moved a constant used as a dividend outside a loop, the check rejected a valid certificate. The repair
also accepts a dividend that the original invariant pins to a literal. 
The repair updated both the certificate checker and its correctness proof. It
left the transform and certificate generator unmodified.

\subsubsection{FMAtarget}
This rejection occurred because of a defect in the certificate generator for the multiply add fusion
transform. Recall that, for each step in the transformed program, the certificate must identify a
corresponding path in the original program. The defect was triggered when a branch's target in the
original program was a multiply fused with the following add into a single
multiply add. The step in the transformed program is the single control transfer from the branch
to the multiply add. The correct corresponding path in the original program therefore ends at the
instruction that the branch targets, specifically the multiply. But the generator extended the path
one instruction too far --- past the multiply to the following add --- so the path incorrectly ended
at the add rather than at the multiply. The certificate checker correctly rejected the certificate.
The repair updated the certificate generator to emit the single instruction path to the multiply.
Note that the correspondence between the multiply followed by add in the original program and the fused
multiply add in the transform is recorded by the step that executes the fused multiply add (not the branch),
with the corresponding path in the original program containing both the multiply and the add instruction.

\subsubsection{Peephole}
When peephole noop removal collapsed a conditional branch so that the branch taken target equals 
the branch fall through, the checker did not distinguish the two branch directions and rejected the resulting degenerate
transition, even though both directions target the same original instruction. The repair
updated the checker to verify both directions of a degenerate branch. 
The repair therefore updated both the certificate checker and its correctness proof. It
left the transform and certificate generator unmodified.

\subsubsection{LICMpreheader}
Loop invariant code motion generated a certificate that incorrectly claimed hoisted constant values hold at
program points a loop bypass path can reach without executing the preheader. The certificate checker
then rejected the certificate. The repair updated the certificate generator to check for this case. 
When the check fires, the certificate generator discards the transform, turns the optimization into a noop,
and generates an identity certificate. Note that the transform
itself is correct but the coding agent chose to discard the optimization rather than update the certificate
generator and certificate checker to handle the case. The optimization effect is the same before and after
the fix (the transform is discarded), but the fix eliminates the certificate rejection. This is essentially
an engineering decision on the part of the coding agent that the benefit of applying the optimization in 
this case does not justify the expenditure of the engineering effort required to update the certificate checker and proofs.

\subsubsection{Unreachable}
The certificate checker validated invariants and transitions at every program point, including points
reachable only through statically unreachable control flow edges. The empty invariants at unreachable 
points cannot discharge the relevant proof obligations. The repair updated the certificate checker to 
compute a reachability set and skip checks of proof obligations at unreachable points. 
The correctness proof establishes that no reachable point is ever skipped. This repair was the largest 
of the campaign. 
The repair updated both the certificate checker and its correctness proof. It
left the transform and certificate generator unmodified.

\subsubsection{RegAlloc}
The register allocator computed variable interference from liveness alone. It then incorrectly
concluded that a variable defined by a
dead store did not interfere with variables live at the dead store. The graph coloring
algorithm then assigned the dead stored variable and a still live variable to the same register, so
the dead store incorrectly overwrote the live value.
The transform was therefore incorrect and the certificate checker correctly rejected the certificate.
The repair corrected the register allocation error by making
each definition interfere with everything live at its program point.

Because the certificate is generated from the coloring, correcting the allocator alone eliminated
the rejection --- the certificate generator and checker are unchanged, and the regenerated certificate now checks. 
Only the transform changed.

\subsubsection{CmpFold}
The expression simplifier in the certificate checker did not 
simplify the operands of integer comparisons before checking for equality. 
It was therefore unable to recognize some cases
when the original and transformed sides of a hoisted comparison were equivalent but
syntactically unequal. 
The repair makes the simplifier recurse into integer comparison operands to simplify the
operands before checking for equality. 
The repair updated both the certificate checker and its correctness proof. It
left the transform and certificate generator unmodified.

\subsubsection{BoolWrap}
The same certificate checker expression simplifier also did not simplify operands
of boolean expressions (negation, conjunction, disjunction, comparison against a literal, and
boolean casts). The defect caused the checker to reject certificates for some hoisted boolean
expressions. The fix is the same as for the CmpFold defect --- make the simplifier recurse
into operands of boolean expressions to simplify the operands before checking for equality. 
The repair updated both the certificate checker and its correctness proof. It
left the transform and certificate generator unmodified.

\subsubsection{FMAtemp}
Recall that certificates state relations between the values of variables in the 
original and transformed programs. The multiply add fusion optimization excluded the
variable that the multiply in the original program wrote (before the add) 
from occurring anywhere in these relations, which is sound only for fresh temporaries that are written only once
(to store the multiply result) and used only once (as an operand of the following add). 
The repair applies the fusion optimization only when these conditions hold. An 
alternate repair would update the certificate generator to generate correct certificates even
for multiply add fusions whose original multiply target is written elsewhere in the program. 
This is another case of an engineering decision on the part of the coding agent that the benefit of applying the optimization in 
this case does not justify the expenditure of the engineering effort required to update the certificate generator. 
Only the transform changed. 

\subsubsection{DAEstore}
Dead assignment elimination generated a certificate that asserted a dead variable evaluated
to a constant in the transformed program. The certificate checker could not verify the 
assertion and rejected the certificate. The repair dropped the dead variable from the relation between the
original and transformed programs, asserting nothing about the variable, 
which is sound because the variable is dead. Only the certificate generator changed. The
transform and certificate checker are unmodified.

\subsubsection{DAEobservable}
Constant propagation creates a loop condition that is always true so that there 
is an infinite loop and the
halt instruction after the loop is unreachable. In this case the certificate generator
asserts nothing about the relation between the values of original and transformed 
variables at the halt. But the checker required the certificate to prove
that observable variables have the same values at 
halt instructions in the original and transformed programs. The checker therefore
rejected the certificate. The repair updated the certificate checker to require
proofs that observable values have the same values only at reachable halt instructions. 
The repair updated both the certificate checker and its correctness proof. It
left the transform and certificate generator unmodified.

In general, the code that causes these rejections and the corresponding repair code is well structured
and the repairs generalize within their intended scope. There is no evidence of reward hacking.

\begin{table*}[t]
  \centering
  \caption{Certificate Checker Repairs (T3b, SPE generator). Columns as in Table~\ref{tab:t3a-repairs}.}
  \label{tab:t3b-repairs}
  \setlength{\tabcolsep}{4pt}
  \footnotesize
  \begin{tabular}{@{}lll cc r rrrr r r@{}}
    \toprule
    Defect & \makecell{Detect\\Phase} & \makecell{Updated\\Component} & \makecell{Def LOC\\($+$/$-$)} & \makecell{Proof LOC\\($+$/$-$)} & \makecell{Time\\(min)} & \makecell{Input\\Tokens} & \makecell{Cache Read\\Tokens} & \makecell{Cache Write\\Tokens} & \makecell{Output\\Tokens} & Cost & \makecell{Tool\\Calls} \\
    \midrule
    DAEobservable & D1 & Checker & $+2/{-}0$ & $+28/{-}22$ & 44.6 & 5{,}312 & 8{,}893{,}257 & 133{,}994 & 58{,}986 & \$7.29 & 91 \\
    \bottomrule
  \end{tabular}
\end{table*}

\subsection{Parser and ASM Printer}

\begin{table*}[t]
  \centering
  \caption{Parser and ASM Printer Repairs. 
Detect Phase: D = found in
    detect phase; U\,(\textit{x}) = unmasked during verification of
    fix~\textit{x}. }
  \label{tab:t2-repairs}
  \setlength{\tabcolsep}{4pt}
  \footnotesize
  \begin{tabular}{@{}lll cc r rrrr r r@{}}
    \toprule
    Defect & \makecell{Detect\\Phase} & \makecell{Updated\\Component} &
      \makecell{Def LOC\\($+$/$-$)} & \makecell{Proof LOC\\($+$/$-$)} &
      \makecell{Time\\(min)} &
      \makecell{Input\\Tokens} & \makecell{Cache Read\\Tokens} &
      \makecell{Cache Write\\Tokens} & \makecell{Output\\Tokens} &
      Cost & \makecell{Tool\\Calls} \\
    \midrule
    VarMultiline & D                  & Parser  & $+2/{-}0$  & $0/0$ & 3.6 & 3{,}737 & 875{,}702     & 36{,}954 &  9{,}320 & \$1.06 & 24 \\
    BraceSyntax  & U\,(VarMultiline) & Printer & $+4/{-}2$  & $0/0$ & 5.7 & 3{,}741 & 918{,}679     & 37{,}270 & 16{,}932 & \$1.28 & 24 \\
    StringEscape & U\,(VarMultiline) & Printer & $+10/{-}2$ & $0/0$ & 3.3 & 3{,}610 & 686{,}873     & 27{,}723 &  8{,}269 & \$0.85 & 23 \\
    BoolLiteral  & U\,(VarMultiline) & Parser  & $+2/{-}2$  & $0/0$ & 3.7 & 3{,}876 & 1{,}267{,}861 & 44{,}256 & 10{,}538 & \$1.36 & 31 \\
    \bottomrule
  \end{tabular}
\end{table*}

The initial 30 minute detect phase processed 13,770,000 programs, all of which
trigger a parse fail defect --- the Axon AST printer was apparently never tested against
the Axon parser. Table~\ref{tab:t2-repairs} presents numbers from the ensuing repair phase.
The phase repaired four defects, one of which was surfaced during the detect phase (D) and three
of which were unmasked during defect repair validation (U). Because every generated program
failed to parse its first line --- the multi-variable declaration --- only that defect is
visible in the detect phase. Repairing that defect lets parsing proceed past the first line and unmasks the other three defects. For two
of the defects the agent chose to repair the Parser, for the other two defects the agent chose to repair the
Printer. The repairs were relatively small and did not involve proof updates. The remaining
statistics characterize the resources required to perform the repairs, all of which landed quickly
with small resource consumption. After the four repairs landed, a fifth session confirmed that a
related round trip mismatch class (boolean literal reprinting) had already been eliminated by the
BoolLiteral repair, and a sixth session surfaced no further defect, for a total repair cost of \$5.38.
When the repair phase finished, the second detect phase processed 18,075,000 programs with no defects surfaced.

\subsubsection{VarMultiline}
The VarMultiline defect occurred because the parser assumed a single var declaration for program variables, while
the printer generated a var declaration per variable. The Parser repair recursively processes the multiple variable declarations.

\subsubsection{BraceSyntax}
The BraceSyntax defect occurred because the 
printer generated keywords the parser did not recognize (then for if statements and do for while statements) and
because it did not include if and while code bodies in braces. The Printer repair dropped the keywords and put the bodies in
braces.

\subsubsection{StringEscape}
The StringEscape defect occurred when the printer generated a raw string that contains characters (for example, newline) that the parser assumes are normally escaped (for example, newline would be \textbackslash n). The Printer
update makes the printer generate normally escaped strings.

\subsubsection{BoolLiteral}
The BoolLiteral defect occurred when the parser generated $0==0$ for a true boolean literal and 
$0 != 0$ for a false boolean literal. Not really a defect, but it produced a different AST program. The Parser 
repair generates true and false. 

In general, the parser and printer code surrounding these defects and repairs is well structured
and the repairs generalize correctly across programs written in the language. There is no evidence of 
reward hacking. 

\comment{
\section{Discussion}

A potential reason for the lack of reward hacking for ASM operational 
semantics and certificate repairs is that they interface with proofs. 
The headline correctness theorems are proved over the operational semantics, 
certificate checker repairs must reprove the certificate checker correctness
theorems, and transform and certificate repairs feed verified certificate
checker. 
}

\section{Threats to Validity}

This research targets the Axon compiler, which compiles a relatively small language. The
findings in this paper may not generalize to compilers that compile existing larger
languages. Given that Axon was produced by a coding agent, we do not necessarily expect
our findings to generalize to existing compilers produced by human developers --- one of the
novel aspects of our research is its application to code produced by a coding
agent, which may differ in significant ways from code produced by human developers. 
The Axon development was driven by a relatively small benchmark set. It is unclear
if our findings will generalize to compilers whose development was driven by a larger
benchmark set or were more extensively tested during development (although the techniques
presented in this paper may help productively drive testing during development). 
Finally, the research focuses on compilers and may not generalize to other kinds of software
systems, which in general have been less intensively studied and are less well understood
in comparison with compilers. This fact could impair the ability of the coding agent to 
reason as effectively about other kinds of software as about compilers. All research was performed using
Claude Code 4.8. Given the rapid development of coding agents the results may change
as newer coding agents with enhanced capabilities are released. Runs with same
coding agent may, because of agent randomization, produce different results. 

\section{Related Work}

We discuss related work in compiler testing and automated defect repair, including proof repair. 

\subsection{Compiler Testing}

CreLLVM updates two LLVM passes to generate credible compilation 
certificates~\cite{kang2018crellvm}. 
Failed certificate checks found four defects (two in each pass). 
Alive2 uses bounded translation validation (implemented by an SMT solver) to attempt
to verify bounded equivalence of LLVM functions before and after transformation
to verify that the after program refines the before 
program~\cite{DBLP:conf/pldi/LopesLHLR21}. 
Working with LLVM unit tests, Alive2 detected 121 refinement violations in LLVM
passes and triggered changes to eliminate ambiguities in the LLVM Language
Reference.  Our research differs in that 1) both the compiler (Axon) and ACDC were written by coding agents, not human developers, enabling an evaluation of coding agent reward hacking, 2) ACDC targets a verified compiler (Axon), 3) unlike CreLLVM, all Axon passes use credible compilation, enabling end to end compiler correctness proofs, 4) unlike 
Alive2 (which uses only bounded SMT checks), the Axon certificate checker comes
with a correctness proof (if the checker accepts, the transformed program 
refines the original program), and 4) ACDC deploys 
agents to automatically generate repairs (CreLLVM and Alive2 report 
defects to human developers). More broadly, the goal of our certificate checker research
is to automatically generate repairs that increase the range of optimizations that 
produce certificates that check ---
the goal of CreLLVM and Alive2 is to find compiler defects for human developers to repair. 

Differential compiler testing compares results from the same 
(typically randomly generated) program 
compiled on different compilers or optimization settings, with any
differences surfacing a defect~\cite{DBLP:journals/pacmpl/LivinskiiBR20,DBLP:conf/pldi/YangCER11}. Because Axon is verified to generate correct code, techniques that rely on 
detecting incorrect compilation can only surface defects in unverified code. 
Here ACDC deploys more targeted techniques that leverage the structure of
the specific unverified component (parser, ASM operational semantics, text printers). 
But ACDC can (and does) reuse random program generation techniques originally developed to 
test unverified compilers to obtain programs
that trigger certificate rejections~\cite{DBLP:conf/pldi/YangCER11,DBLP:conf/pldi/ZhangSS17,DBLP:conf/pldi/LeAS14,
DBLP:conf/oopsla/LeSS15, DBLP:conf/oopsla/SunLS16}.
And unlike traditional compiler testing
systems, which target unverified compilers written by human developers, 
ACDC targets a verified compiler generated by a coding agent. 

Two papers discuss the unverified parts of the CompCert~\cite{leroy2009compcert} verified 
compiler~\cite{DBLP:conf/esop/MonniauxB22,DBLP:conf/tap/MonniauxGBL23}, which are known
to contain defects~\cite{DBLP:conf/pldi/ZhangSS17}. While these papers
identify some of the same unverified components as Axon (text interfaces, formal operational
semantics), they do not discuss testing techniques (like those in ACDC) that exploit the compiler structure
to specifically target these components. 

\subsection{Program and Proof Repair} 
Recent agentic program repair research~\cite{DBLP:conf/icse/BouzeniaDP25,DBLP:conf/issta/0002RFR24} provides LLM-based
agents with a variety of general tools (reading lines of code, searching
a code base, traditional fault localization), that the LLM then flexibly
deploys
to generate patches for defects identified in bug reports, with testing
used to guide patch acceptability. Agentless, in contrast, follows
a simple localization, repair, and patch validation process~\cite{DBLP:journals/pacmse/XiaDDZ25}. Despite the different approaches, these systems are designed
to repair defects in software systems generally and do not specifically
leverage structure present in a more focused class of software systems. 
ACDC, in contrast, leverages the structure of the 
Axon compiler to deploy targeted
testing techniques whose defect and rejection reports help the repair
agent precisely identify relevant code to repair. ACDC also leverages the 
Axon compiler structure to write focused prompts that precisely guide
the repair process based on the type of defect or rejection. 
And in addition to repairing code, Axon also updates proofs of 
program correctness. 

Classical program repair research~\cite{repair-living-review} 
typically deploys search algorithms 
(often in combination with machine learning) to generate
small focused patches. ACDC combines the 
capabilities of modern coding agents with defect and certificate
rejection records produced by targeted testing
techniques to generate repairs with much larger
scope, including updating correctness proofs when proof relevant code
is modified. 

The goal of previous proof repair research 
is to repair proofs, for example in response to proof changes such as
theorem and tactic changes or changing types over which the proofs 
operate or because a generated proof is 
incorrect~\cite{DBLP:conf/cpp/RingerYLG18,DBLP:conf/sigsoft/FirstRRB23,DBLP:journals/pacmpl/ViolaFR25}. 
Given an original and evolved version of an OCaml program, 
Sisyphus implements mostly-automated proof repair that relies
on a human developer to fill in key missing proof elements~\cite{DBLP:journals/pacmpl/GopinathanKS23}. 
ACDC, in contrast, automatically updates proofs of program correctness in response to 
repairs that update the operational semantics or source code of the program
(for ACDC, the program is the compiler), 
including changes to verified code (the certificate checker).
We are aware of no previous work in which automatic code repairs trigger 
automatic updates of relevant proofs.

\section{Conclusion}
\label{sec:conclusion}

Verification and checked computations can be effective mechanisms for ensuring
compiler correctness specifically and the correctness of code produced by coding agents more generally. 
Despite these techniques, even extensively verified systems typically contain some unverified
code. We present ACDC, a system that targets the structure of the Axon verified compiler to 
deploy specialized testing techniques for the components in this compiler. ACDC targets unverified
text interfaces, operational semantics, checked but not verified optimizations, and the
verified certificate checker, with the 
resulting testing information providing the repair agents with targeting information
they can use to localize, diagnose, and repair surfaced defects and certificate rejections. 
The experimental results highlight the impressive effectiveness of the repair agents at
repairing both code and correctness proofs. Because both Axon and ACDC were developed
solely by (supervised) coding agents, reward hacking may be a potential concern. Despite
evidence of reward hacking in other contexts, we found no evidence of reward
hacking in either the Axon compiler or the generated repairs. 

\section*{Data Availability}
\label{sec:data-availability}

The code for this paper is publicly available at
\url{https://github.com/rinard/AxonTestAndRepair}. It contains the baseline Axon
compiler, the compiler after each repair campaign, 
the ACDC harness that runs the testing and repair processes, 
each surfaced defect and certificate rejection in
order of discovery, code diffs for each repair, repair session transcripts, and the scripts and extracted
data. See the \texttt{README.md} and \texttt{BUILD.md} files at the top level of the repository.


\bibliographystyle{IEEEtran}
\bibliography{references}

\end{document}